# Helical fluctuations in the Raman response of the topological insulator Bi$_2$Se$_3$

V. Gnezdilov,[1] Yu. G. Pashkevich,[2] H. Berger,[3] E. Pomjakushina[4], K. Conder[4], P. Lemmens[5]

[1]*B. I. Verkin Institute for Low Temperature Physics and Engineering of the National Academy of Sciences of Ukraine, Kharkov 61103, Ukraine;*

[2]*A. A. Galkin Donetsk Phystech of the National Academy of Sciences of Ukraine, Donetsk 83114, Ukraine;*

[3]*Institute de Physique de la Matiere Complexe, EPFL, CH-1015 Lausanne, Switzerland;*

[4]*Laboratory for Development and Methods, Paul Scherrer Institut, CH-5232 Villigen, Switzerland;*

[5]*Institute for Condensed Matter Physics, TU Braunschweig, D-38106 Braunschweig, Germany.*

The topological insulator Bi$_2$Se$_3$ shows a Raman scattering response related to topologically protected surface states amplified by a resonant interband transition. Most significantly this signal has a characteristic Lorentzian lineshape and spin-helical symmetry due to collision dominated scattering of Dirac states at the Fermi level $E_F$ on bulk valence states. Its resonance energy, temperature and doping dependence points to a high selectivity of this process. Its scattering rate ($\Gamma \approx 40$ cm$^{-1} \approx 5$ meV) is comparable to earlier observations, e.g. in spin-polaron systems. Although the observation of topological surface states in Raman scattering is limited to resonance conditions, this study represents a quite clean case which is not polluted by symmetry forbidden contributions from the bulk.

PACS: 73.43.Lp, 73.20.At, 63.22.-m

## *Introduction*

Topological insulators (TI) [1-5] realize a novel quantum state of matter with a topologically protected, conducting surface state on a semiconductor with a gap due to strong spin-orbit coupling [6,7]. This state opens perspectives for applications [8] as well as for fundamental issues of quantum electrodynamics, such as Majorana fermions, magnetic monopoles and axions [9,10]. The compound $Bi_2Se_3$ [11,12] is a seemingly simple realization of a 3D topological insulator with a narrow bulk gap ($\Delta$=0.3 eV) and one relativistic Dirac cone of these surface states [6,7]. The observation of charge helicity, caused by strict spin-momentum locking, and other interesting properties of TI's, is limited to a small number of experimental techniques, in particular ARPES, as in other probes often bulk contributions dominate. Interesting questions to be investigated relate to the effect of doping/defects, warping of the Fermi surface, coupled modes and collective/exotic excitations [1-5].

Here we report the first Raman observation of helical, conducting surface states on 3D topological insulators. The 2D spin-textured metal shows up as a continuum in a helical scattering channel with a decisive energy and temperature dependence. Its Lorentzian lineshape resembles observations in low-dimensional correlated electron systems and points to enhanced electronic correlations at the Dirac point. Raman scattering (RS) is a powerful tool in the study of electronic properties of solids and has been widely applied to quasi-2D electron systems [13]. We believe that this technique will provide additional and relevant information about the unique topological surface states that is complementary to spin and angle resolved photoemission spectroscopy.

## *Experimental*

Single crystals of $Cu_xBi_2Se_3$ were grown using the Bridgman method. Cu doping with x=0, 0.07 and 0.2 is due to singly ionized interstitial Cu atoms [14]. Freshly-cleaved crystals with flat shiny *ab* surfaces and stabilized surface electron concentration [7] were immediately transferred into a cryostat (RT-2.8 K). RS measurements were performed in quasi-backscattering geometry. We used the excitation lines $\lambda$=632.8 nm and $\lambda$=532.1 nm with a power of less than 3 mW. The setup allowed identifying intrinsic low-frequency scattering for frequencies above ~12 cm$^{-1}$. The spectra were obtained with the incident ($e_i$) and scattered ($e_s$) light polarized in various configurations, including ($e_i$,$e_s$) = (***x***,***x***), (***y***,***x***), (***R***,***R***), (***L***,***L***), (***R***,***L***), and (***L***,***R***), where *x/y* directions lie in the (001) plane, *R/L* are right/left circular polarizations.

$Bi_2Se_3$ has a rhombohedral crystal structure (space group $R\bar{3}m = D_{3d}^5$, N 166 with one formula unit in the primitive cell). The site symmetry 2*c* and 1*a* of Bi/Se(1) and Se(2) [12], respectively, leads to $2A_{1g} + 2E_g$ Raman- and $2A_{2u} + 2E_u$ infrared-active phonon modes [11]. Previous RS studies have investigated single crystals [11,15], thin films, ultrathin nanoplates and exfoliated crystal sheets [16].

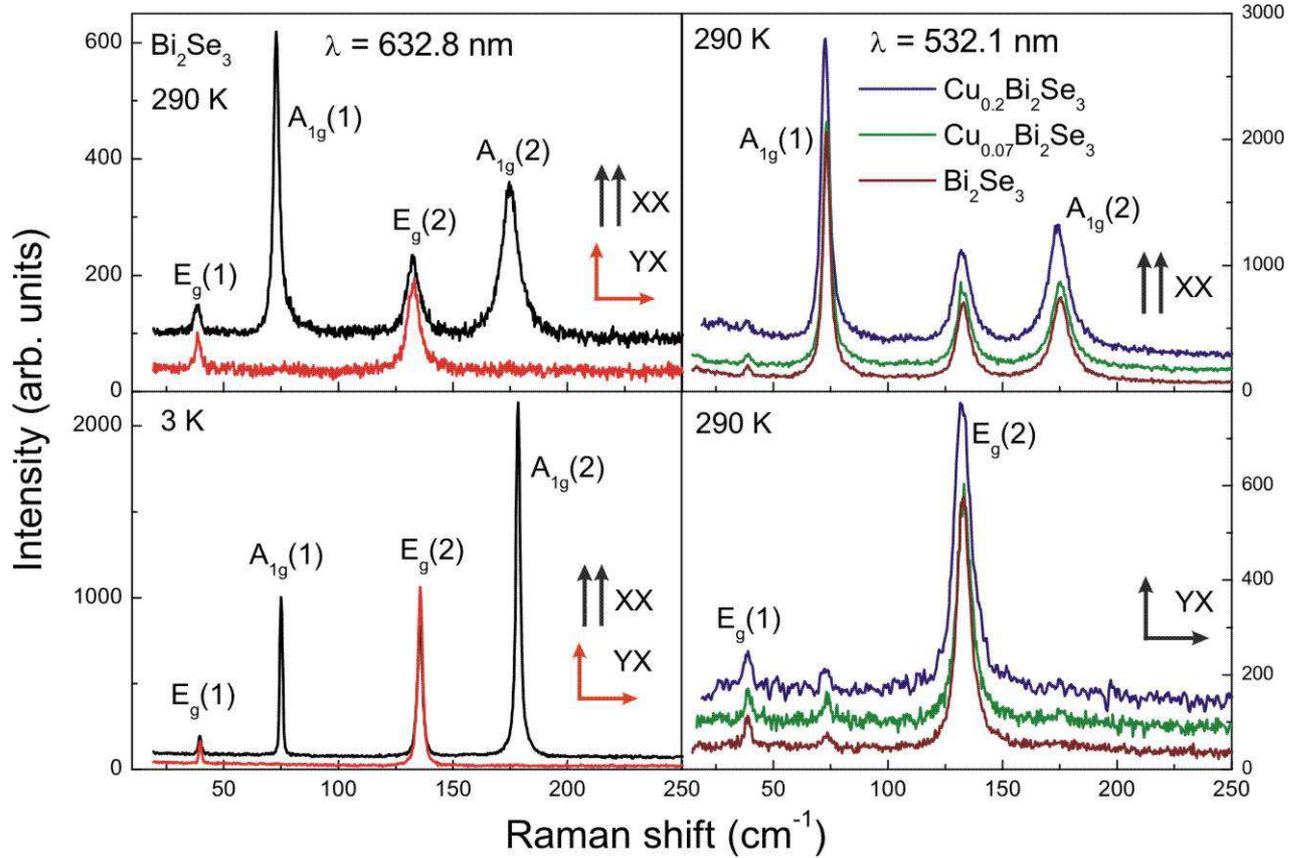

Fig. 1: Raman spectra of $Cu_xBi_2Se_3$ in different polarizations, excitation energies and doping to demonstrate the high symmetry specificity. In nonresonance or at high temperatures only the 4 Raman allowed phonon modes are observed. Left panel) data for $Bi_2Se_3$ with 632.8-nm excitation at 290K and 3K. Right panel) 290K data comparing different doping levels.

## *Results*

Figure 1 (right panel) presents Raman spectra with two $E_g$ and two $A_{1g}$ phonons for $\lambda = 632.8$ nm excitation at 295 K and 3 K, see Table 1 for phonon frequencies and results of calculations [17]. The modes can be easily distinguished due to the off-diagonal Raman tensor components of the $E_g$ modes that are projected in measurements in *YX* polarization. Noteworthy, the $A_{1g}$ modes have generally a larger intensity than the $E_g$ modes. Our data are in good agreement with earlier experiments. [11,15,16]. There exist no phonon anomalies with decreasing temperatures, only moderate phonon frequency shifts and decreasing linewidths are observed. The phonon intensities increase moderately changing the excitation line to $\lambda = 532.1$ nm due to resonance effects, see right panel of Fig. 1. At T=290 K there is no big change of the phonon intensity with Cu doping. The phonon lines remain sharp for $x < 0.2$ and the spectra show no background. In general the reproducibility of the Raman spectra is very high.

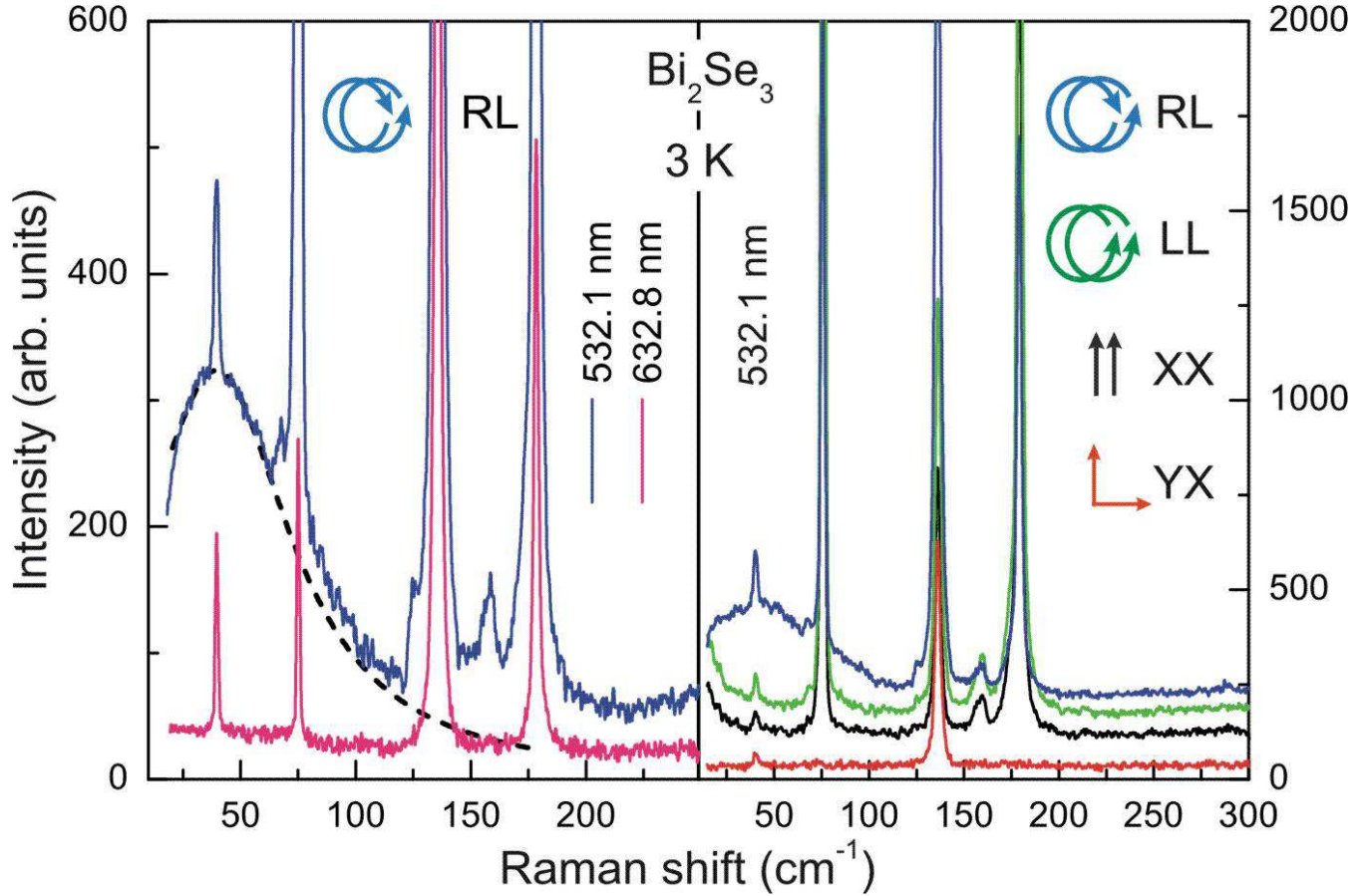

Fig. 2: Strong polarization and resonance dependence of a broad continuum around 40 cm$^{-1}$ at T=3K. Left panel) helical scattering component (*RL*) with 532.1-nm and 632.8-nm laser excitation. The dashed line shows a fit to a model based on electronic scattering with a single scattering rate. Right panel) Comparison of different scattering geometries with 532.1-nm excitation to prove the helical nature of the continuum.

At low temperatures and with λ=532.1-nm excitation there is a considerable change of the Raman spectra. The *RL* polarization component shows a pronounced maximum, see Fig. 2, left panel. There is good agreement to a fit using a Lorentzian centered at $E_{max}$=39 cm$^{-1}$ with a width (FWHM) of w=80 cm$^{-1}$. This value is much larger than that the observed phonons (w=2.5 cm$^{-1}$) or their dispersion. Therefore the continuum is not attributed phonon scattering. Instead, we attribute it to scattering on electronic surface states enhanced by electronic resonance effects. This is so far the only observation of a dominant helical surface contribution from a topological insulator in a spectroscopy different from ARPES.

Furthermore, in other polarizations quasielastic scattering and sharp peaks in proximity to the phonon modes are noted. The quasielastic mode is evidence for further low energy excitations or coupled modes. The additional phonons are identical in frequency with polar modes known from infra red absorption [11]. They have $A_{2u}$ and $E_u$ symmetry, see Table 1. In the bulk of $Bi_2Se_3$ the inversion center prohibits the observation of these polar modes. However, surfaces and internal interfaces break inversion symmetry. Our observation is consistent with the (001) surface of the single crystal which includes a threefold rotational symmetry of the $Z$-axis and three planes of mirror symmetry. It reduces the rotational symmetry from $D_{3d}$ to $C_{3v}$. In this case two $E_u$ modes that remain degenerate become Raman active as $E$ modes and two former $A_{2u}$ modes become fully symmetric $A$ modes. We have indeed observed $4A+4E$ modes, see Table 1.

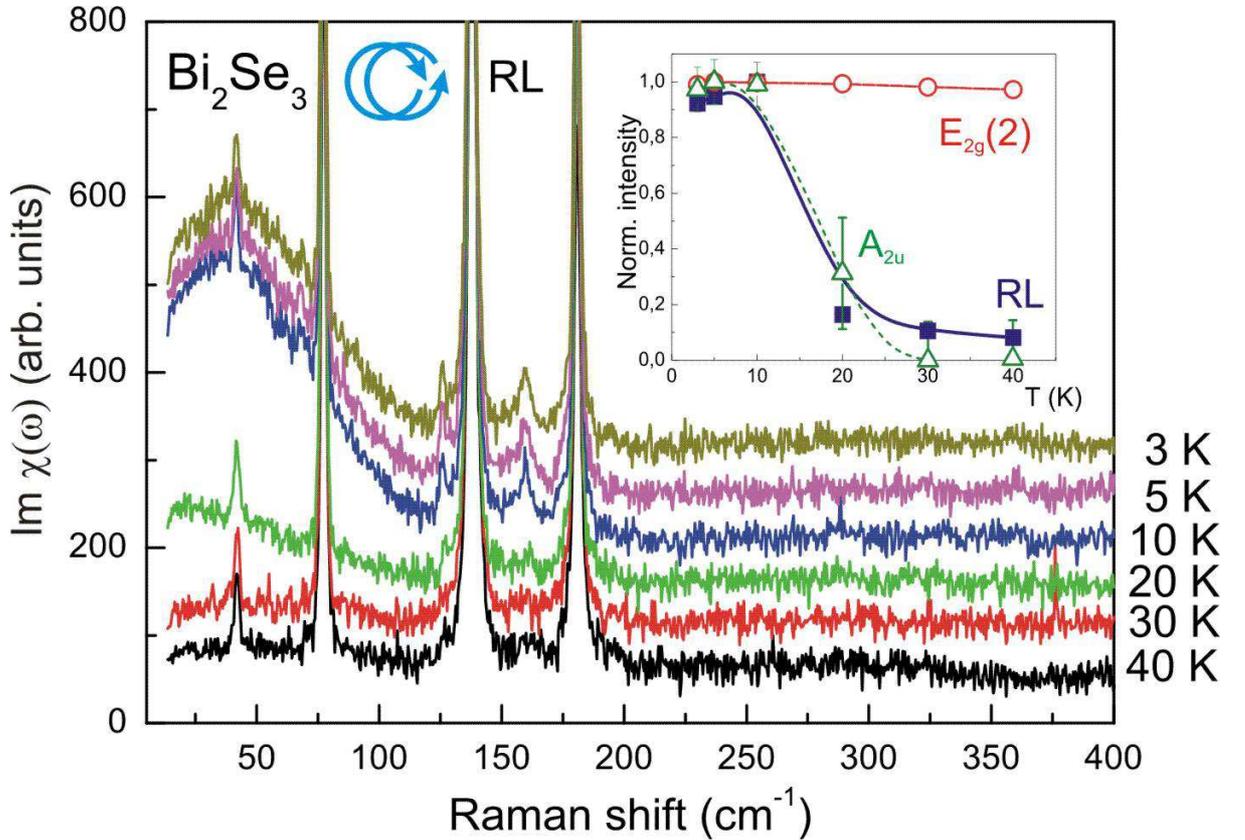

Fig. 3: Pronounced temperature dependence of the helical scattering continuum and the polar phonon modes with 532.1-nm excitation. The Bose corrected Raman response is given an offset for clarity. The inset shows the integrated intensity of the continuum (full squares) compared to the Raman active phonon $E_g(2)$ (open circles) and the polar phonon $A_{2u}(2)$ (open triangles) at 160 cm$^{-1}$.

The additional modes exist only at low temperatures, with a characteristic temperature $T^* = 10$ K, as shown in the inset of Fig. 3. In contrast, the modes that are also observed in non-resonance conditions are constant in intensity, see inset. The simultaneous evolution with temperature of the polar phonon mode and the continuum's intensity point to a similar mechanism of enhancement.

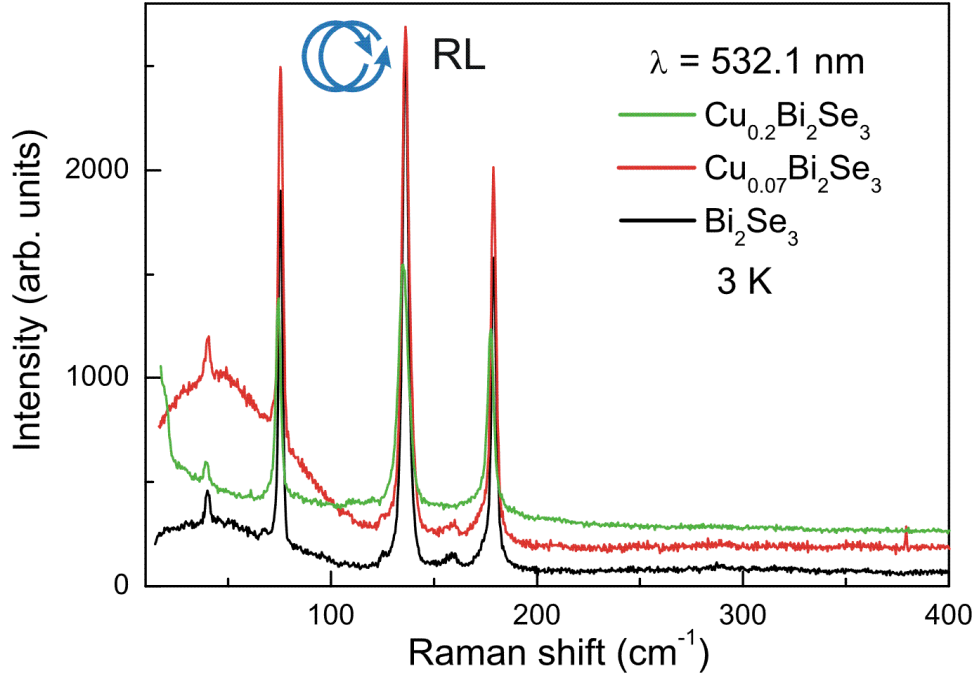

Fig. 4: The helical scattering continuum is amplified with weak doping, $x=0.07$, and suppressed for larger $x=0.2$ in $Cu_xBi_2Se_3$.

In Fig. 4 we show data on the doping dependence of the continuum. The continuum is enhanced for $x=0.07$ compared to $x=0$, without shifting in energy. The other modes do not change their appearance. However, for larger doping the continuum and the polar modes are smeared out or suppressed.

## *Discussion*

In the following we will discuss symmetry aspects of the scattering continuum and the polar surface modes. Both scattering contributions are forbidden in the symmetry of the bulk ($D_{3d}$). The *RL* symmetry component transforms like $(XX - YY) + i(XY + YX)$. Therefore the symmetry allowed $E_g$ modes are both observed in *RL* as well as in *YX* geometry. The scattering continuum, however, is the only signal observed purely in *RL* geometry. It is therefore attributed to the helical surface states. The surface is considered as it is the only chiral medium involved in the scattering process. As we do not see any continuum in the other polarizations, we conclude that only the helical states on the Dirac cone have the required properties to lead to electronic Raman scattering. This is consistent with a very high specificity of these states.

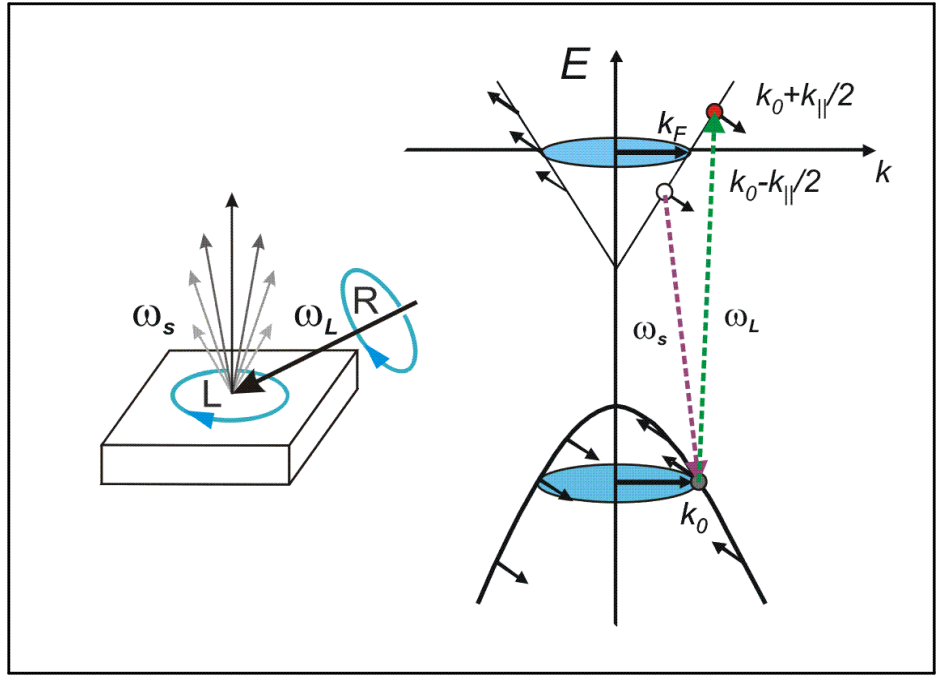

Fig. 5: left panel) Scattering geometry with incident and scattered circular polarized light. Due to the dipole characteristics of the emission following the Raman process the scattered light distributes with different intensities and momenta as symbolized by the arrows. Right panel) Resonant excitation process starting from a bulk valence band state with $k_0$ to a surface Dirac cone state near $k_F$. Fluctuations of the particle-hole continuum on the Dirac cone close to the Fermi energy are probed in resonance based on the momenta estimated from the geometry and the known dispersions [7].

From the presented data we conclude that two consecutive processes are involved in the observation of the continuum: A resonant optical interband transition leads to the selectivity of the involved electronic states with respect to symmetry and excitation energy. The coupling of the electron-hole pairs to electronic fluctuations of the topological states leads to the low energy dynamics, i.e. the line shape and temperature dependence of the continuum. The latter process is specific for RS and its result resembles to systems in the collision dominated regime.

The interband transition from bulk valence band states to the Dirac cone of surface states has a small but well defined wave vector transfer that can be used to estimate the energy range of the involved fluctuations, see left panel of Fig. 5. With an angle of approximately 12° of the incident light to the surface plane our scattering geometry leads to $k_\parallel \sim 0.43 * k_L$, with $k_L$ the wave vector of the incident radiation with $\lambda=532.1$ nm. With quasi particles on the Dirac cone of $v_F =$

$5 \cdot 10^5$ m/sec [18] we achieve a fluctuation energy of $\Delta\omega$=13.4 cm$^{-1}$ fitting well to the onset of the scattering continuum. If we consider spin–plasmon excitations with a square root dispersion we achieve only roughly $\Delta\omega$=1 cm$^{-1}$. So they may not be of relevance here.

The small wave vector around the Dirac cone ($k_F \approx 0.1 Å^{-1}$) confines the initial state to only one bulk valence band from which the $hv_{green}$ =2.33 eV quanta can reach the Dirac cone, Fig. 5, right panel. This energy interval fits perfectly to the experimentally known dispersions in Bi$_2$Se$_3$ [7]. The very steep negative dispersion of this band, the positive dispersion of the Dirac states and dipole selection rules lead to the observed high selectivity of the interband transition. Interestingly, the electron-hole pairs of this interband transition may induce self-doping with an effective redistribution of electron density in the surface layers. This leads to an electric dipole moment perpendicular to the surface and $C_{3v}$ symmetry as a part of the system ground state.

Electronic fluctuations are rarely observed in RS and usually restricted to some correlated electron systems [19-21]. The efficient screening of charge density fluctuations and long range Coulomb interactions can be suppressed by strong scattering involving electron-electron interaction [21] and nesting [22]. In the collision-dominated regime with a single relaxation rate $\Gamma$ the scattering cross section is described by a Lorentzian line with a maximum at $\Gamma$ [21]. Exactly this lineshape is shown in Fig. 2, left panel, with $\Gamma$=39 cm$^{-1}$ and w=80 cm$^{-1}$ [23]. The magnitude of the relaxation rate $\Gamma$ is quite similar to observations in spin-polaron systems [20]. The linewidth w is reduced by approximately a factor of two compared to a model with isotropic scattering [21]. We suggest that these moderate deviations stem from the topological protection that the Dirac states are exposed to. Topological protection suppresses backscattering [24]. However it does not suppress anisotropic scattering with smaller momenta between Dirac surface and bulk states.

In a recent scanning tunneling microscope experiment (STM) on Bi$_2$Se$_3$ an anisotropic interference pattern has been observed at Bi excess surface atoms. This interference shows typical scattering vectors k $\approx$ 0.3 A$^{-1}$ which connect energetically degenerate Dirac surface and bulk states along the $\Gamma$-M direction. No scattering along the G-K direction is observed as in this direction the bulk valence band is lower in energy [25]. The RS continuum in Bi$_2$Se$_3$ is thereby a direct measure of the anisotropic electronic correlations of the carriers at the Fermi energy. Although the general implications of this effect are not totally clear it shows that the electronic properties of these compounds are more complex than previously considered.

Noteworthy are the large variations of the RS continuum with doping and temperature. The initial enhanced and following depletion of the continuum with composition resembles a critical behavior [26-28]. In first approximation it can be ascribed to a shift of the Fermi energy with respect to the bulk states, thereby modifying the scattering vectors. In relation with a collision dominated regime in graphene also drastic effects of temperature have been discussed [26]. However, experimentally a temperature dependence of the topological states in $Bi_2Se_3$ has neither been observed in ARPES [7] nor in STM experiments [25]. Therefore we attribute it to the RS process. Most probably thermally excited carriers induce screening and reduce the electron-electron interactions [27]. Both effects drastically reduce electronic RS.

There exists a general correspondence between the electronic Raman response and the optical conductivity $\sigma(\Omega)$ given by $\chi''(\Omega) \propto \Omega\sigma(\Omega)$ [30]. Using our data the deduced optical conductivity $\sigma(\Omega)$ in the energy range 15-125 cm$^{-1}$ does not show a Drude-like $\sigma(\Omega) \approx 1/\Omega$ response which is expected for spin-plasmon excitations [18,28]. Thus we conclude that we are in a regime far from spin-plasmon excitations and the optical conductivity is related to a 2D spin-textured helical metal with algebraic correlations [22,29].

Finally we will discuss phonons and possible coupled modes. Two of the four polar surface phonon modes observed in resonance have a dynamical polar moment perpendicular to the sample surface ($P_z \parallel$ Z-axis). This could have some implication for the topological states. Coupled electronic/phononic modes have been discussed, e.g. for the case of graphene [30]. For $Bi_2Se_3$ they still await an observation. We have no direct evidence for such a mode. However, the quasielastic scattering observed in *XX* and *LL* polarization could be a candidate for a mixed electronic-acoustic excitation that should motivate further studies at lower energies.

## *Summary*

Besides from photoemission it has been difficult to find a direct manifestation of surface conductive states in other experiments as they are masked by bulk contributions. In the present RS experiment the strong selectivity of both the symmetry constraint and the electronic resonance allows observing a genuine electronic surface contribution. We interpret the continuum as due to fluctuations of Dirac states in the collision dominated regime and its strong temperature dependence to thermally induced carriers that reestablish screening. Our observations demonstrate the versatility of resonance Raman scattering, complementary to ARPES data, for the study of topological protected surface states.

## *Acknowledgement*

This work was supported by DFG and FRSF of Ukraine Grant No. F29.1/014.

TABLE 1. Experimentally observed phonon frequencies (cm$^{-1}$) and ab-initio calculations [17] including spin-orbit interactions for $Bi_2Se_3$ at T=3 K. Modes (*) are attributed to resonance and surface contributions. The $A_{2u}(1)$ mode is observed as a weak shoulder of $E_g(2)$. Modes (**) are from FIR at T=15 K [11].

| Modes symmetry | Frequencies (cm$^{-1}$) | |
|---|---|---|
| | Ab-inito calculations [17] | Experimentally observed in RS |
| $E_g(1)$ | 41.49 | 38.9 |
| $A_{1g}(1)$ | 75.42 | 73.3 |
| $E_g(2)$ | 137.06 | 132.9 |
| $A_{1g}(2)$ | 171.02 | 175.4 |
| $E_u(1)$ | 80.29 (61**) | 68* |
| $E_u(2)$ | 130.52 (134**) | 125* |
| $A_{2u}(1)$ | 137.29 | 129* |
| $A_{2u}(2)$ | 161.35 | 160* |